\documentclass[12pt]{article}
\usepackage{url}
\input{epsf}

\font\bm=cmmib10 at 10pt
\font\bms=cmmib10 at 7pt \textfont9=\bm \scriptfont9=\bms
\usepackage{amsfonts}
\usepackage{multirow}
\mathchardef\balpha= "790B
\mathchardef\bbeta= "790C
\mathchardef\bTheta= "7902
\mathchardef\bzeta= "7910
\mathchardef\bOmega= "790A
\mathchardef\bGamma= "7900
\mathchardef\bDelta= "7901
\mathchardef\bPhi= "7908
\mathchardef\bphi= "791E
\mathchardef\bomega= "7921
\mathchardef\bxi= "7918
\mathchardef\bet= "7911
\mathchardef\brho= "791A
\mathchardef\btau= "791C
\mathchardef\bmu= "7916
\mathchardef\bvarpi= "7924

\def \lvec{(\kern-.26em(}
\def\pmb#1{\setbox0=\hbox{#1}%
	\def \lvec{(\kern-.26em(}
	\kern-.025em\copy0\kern-\wd0
	\kern.05em\copy0\kern-\wd0
	\kern-.025em\raise.0433em\box0 }
\mathchardef\btheta= "7912
\usepackage{amsmath}
\usepackage{authblk}

\usepackage{graphicx}
\usepackage{dcolumn}

\begin{document}
\title{Atmosphere and Greenhouse Gas Primer}
\author[1] {W. A. van Wijngaarden}
\author[2] {W. Happer}
\affil[1]{Department of Physics and Astronomy, York University, Canada}
\affil[2]{Department of Physics, Princeton University, USA}
\renewcommand\Affilfont{\itshape\small}
\date{\today}
\maketitle

\begin{abstract}
\noindent 
We discuss the basic ways greenhouse gases affect radiation transfer in Earth’s atmosphere. We explain how greenhouse gases like water vapor, H$_2$O, or carbon dioxide, CO$_2$, differ from non-greenhouse gases like nitrogen, N$_2$, or oxygen, O$_2$.  Using simple thermodynamics and fluid mechanics, we show that the atmosphere of a planet with sufficiently high concentrations of greenhouse gases must develop a convecting troposphere between the surface and the tropopause altitude. The planet must also develop a non-convecting stratosphere for altitudes above the tropopause.  In the simplest approximation of an atmosphere that is transparent to sunlight and has frequency-independent opacity for thermal radiation (an infrared gray atmosphere), one can find simple formulas for the tropopause altitude, and for the altitude profiles of pressure and temperature.  The troposphere is nearly isentropic and the stratosphere is nearly isothermal. The real atmosphere of the Earth is much more complicated than the simple model, but it does have a  troposphere and a stratosphere.  Between the surface and the tropopause the entropy per kilogram of real tropospheric air increases slowly with altitude. The entropy increases much more rapidly with altitude in the stratosphere.  The stratosphere has a nearly isothermal lower part and a hotter upper part due to absorption of solar ultraviolet radiation by ozone.  The thermal opacity of the real atmosphere has a complicated frequency dependence due to the hundreds of thousands of vibration-rotation transitions of its greenhouse molecules. Unlike the simple model where nearly all radiation to space originates at the tropopause altitude, radiation to space from Earth’s real atmosphere originates from both the surface and all altitudes in the troposphere. A small additional amount of radiation originates in the stratosphere. When these complications are taken into account, model calculations of the thermal radiation spectrum at the top of the atmosphere can hardly be distinguished from those observed from satellites.
\end{abstract}
%
%
\newpage
\section{Introduction}
Worldwide industrialization and the associated combustion of fossil fuels have increased the concentrations of carbon dioxide (CO$_2$) and methane (CH$_4$) since 1750.  These gases along with nitrous oxide (N$_2$O) and assorted lesser players like halocarbon refrigerants are examples of “greenhouse gases”.  It should be noted that by far the most abundant greenhouse gas in the atmosphere is water vapor.  There is little that one can do about water vapor on our watery planet Earth, with 70\% of its surface covered by oceans.

Greenhouse gases were first discovered by John Tyndall in the course of brilliant experimental work in the 1850’s \cite{Tyndall}. Tyndall recognized that greenhouse gases warm Earth’s surface.  Some 50 years later Svante Arrhenius made the first theoretical estimates of how much surface warming would result if atmospheric concentrations of carbon dioxide were doubled \cite{Arrhenius}. The atmosphere and oceans are so complicated that to this day no one knows what the exact warming will be. But basic physics and the geological record indicate that the warming will be small and probably good for life on Earth.  The additional carbon dioxide of the past century has already benefitted agriculture, forestry and photosynthetic life in general \cite{Idso,Zhu}.  Both Tyndall and Arrhenius thought that greenhouse warming was a good thing.

\section{Earth's Atmosphere}
We begin with the basic physics of heat transfer in Earth’s atmosphere. Recall that the largest part of Earth’s atmosphere is a gas, a state of matter where molecules or atoms spend most of their time hurtling through free space. The molecules have occasional collisions with others. At sea level each molecule experiences a collision about once a nanosecond (a billion collisions per second). A collision sends a colliding pair of molecules careening off in different directions, like the collisions of billiard balls on a pool table. If the molecule consists of two or more atoms, like most molecules of Earth’s atmosphere, some of the energy of internal vibrations and rotations can be exchanged.  But the sum of the translational, rotational and vibrational energies of a colliding pair of molecules is the same before and after the collision. The total energy of the colliding pair is conserved, even though the energy of one of the colliding molecules may be bigger and that of the other smaller after the collision.

If one were to follow an individual molecule over the billion or so collisions it experiences every second, it would turn out that the molecule has a well-defined probability to be found in any vibration-rotation state of quantized energy $E_j$. This is an example of the famous ergodic theorem of statistical physics, and it describes a system in local thermodynamic equilibrium. The probability of finding the molecule in a state of energy $E_j$ is proportional to the Boltzmann factor, $e^{-E_j/k_BT}$.  Here $T$ is the absolute temperature of the gas, and Boltzmann’s constant has the value

\begin{equation}
k_B = 1.38 \times 10^{-23} {\rm J/K}. 					
\label{eqn1}
\end{equation}

\noindent Austrian Ludwig Boltzmann, for whom Boltzmann’s constant $k_B$ is named, and his thesis advisor, Joseph Stefan, played very important roles in our understanding of thermal radiation, statistical mechanics and thermodynamics. 

In addition to molecules, Earth’s atmosphere also contains radiation. Sunlight is present during the day. But thermal radiation is present both day and night.  Like a hot coal from a campfire, Earth glows in the dark with thermal infrared radiation. There is much more thermal radiation near the warm equator than near the colder poles or near the surface than at the cold tropopause, around 11 km in altitude for temperate latitudes. 

It is often convenient to think of the radiation as photons, which are a bit like air molecules, and which have quantized energies $E = h\nu$, where $h$ is Planck’s constant and $\nu$ is the photon frequency.  Inside a dense, nighttime cloud, the radiation can be very close to thermal equilibrium with the air molecules and cloud particulates, all of which will have nearly the same temperature. Then the radiation is well described by Planck’s formula for blackbody radiation, the very first formula of quantum physics. But most of the time, atmospheric radiation is far from thermodynamic equilibrium because the photon mean free paths are much longer than the length scales for atmospheric temperature changes.

The energy density of radiation in Earth’s atmosphere is many orders of magnitude smaller than the translational, rotational and vibrational energy densities of the molecules. Radiation has relatively little heat capacity. It is ludicrous to hear about “heat trapped” as photons.  Molecules account for almost all of atmospheric heat.

A substantial fraction of thermal-radiation frequencies is in the “infrared window” where there is negligible clear-sky opacity between the surface and outer space.  Here, there is extremely efficient “ballistic” heat transfer by radiation.

Greenhouse gases can efficiently absorb thermal radiation with frequencies in opaque spectral regions. But greenhouse gases almost never scatter thermal radiation.  This is unlike radiation transport of visible light through clouds or the Rayleigh scattering of blue and ultraviolet light that makes the blue sky of a sunny day. Visible photons can be scattered hundreds of times with little absorption.

Although greenhouse gases do not scatter thermal radiation, they do emit it spontaneously.  So thermal radiation in opaque spectral regions is absorbed and independently reemitted as though it were being isotropically scattered. The kinetic theory of gases tells us that the diffusion coefficient $D$ of a gas of molecules (or of a gas of photons) of mean free path $l$ and velocity $v$, is $D = vl/3$.  For quasi-diffusional transport of thermal radiation in opaque spectral regions, the mean free paths, $l$, of photons between absorptions and reemissions are much longer than those of molecules. This greatly increases the heat-transport efficiency of “diffusing” photons, compared to that of diffusing molecules. Photons also move at the speed of light $c$, about a million times faster than the fluctuating molecular speed $v$, with an average value close to the speed of sound in air. So both $l$ and $v$ are orders of magnitude larger for photons than for molecules. 

In summary heat transport by thermal radiation in Earth’s atmosphere is orders of magnitude faster than heat transfer by molecular diffusion.  Heat transfer by conduction in air (that is, by molecular diffusion) is so small that it is normally irrelevant compared to heat transfer by radiation or heat transport by convection. Heat convection by moist air, which can carry lots of latent heat, as well as sensible heat, is especially important.

\subsection{Greenhouse Gases}

What is a greenhouse gas? It is a gas that is almost transparent to sunlight, and allows the Sun to heat the ground on a cloud-free day.  But greenhouse gases have significant opacity for the thermal radiation that is constantly emitted by Earth’s surface and by the warm greenhouse gases themselves. The nitrogen molecules, N$_2$, and oxygen molecules, O$_2$, which make up 99\% of Earth’s atmosphere, are not greenhouse gases, since they are nearly transparent to both sunlight and thermal radiation.  

Vibrating or rotating electric dipole moments are the most efficient molecular “antennas” for emitting or absorbing electromagnetic radiation. The electric dipole moment of a molecule is the product of the spatial separation between the center of negative electron charge and the center of positive nuclear charge, and the magnitude of the positive charge. By symmetry, the centers of both positive and negative charges of N$_2$ and O$_2$ are at the center of the molecule (and at the center of mass). There is no separation of the charge centers, so the dipole moment is zero, no matter how the molecules vibrate or rotate. The diatomic molecule carbon monoxide, CO, has much the same electronic structure as the nitrogen molecule N$_2$, but because of the lack of symmetry, the O end of the CO molecule has a negative charge and the C end has a positive charge. Therefore, CO is a greenhouse gas, but isoelectronic N$_2$ is not.  The greenhouse effects of CO are so much smaller than those of H$_2$O and CO$_2$ that CO gets little attention in discussions of climate.

\begin{figure}[t]
\begin{center}
\includegraphics[height=100mm,width=.6\columnwidth]{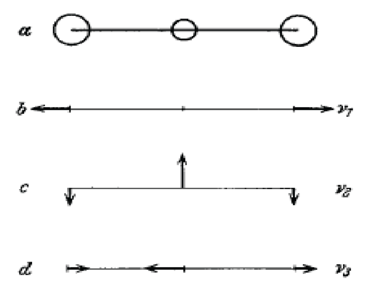}
\caption{The three vibrational modes of the CO$_2$ from Fermi's classic paper on mode mixing \cite{Fermi}.  For the most abundant isotopolgue, $^{16}$O $^{12}$C $^{16}$O, the mode frequencies are:  symmetric stretch (b) $\nu_1 = 1388$ cm$^{-1}$; (c) bending $\nu_2= 667$ cm$^{-1}$; and (d) asymmetric stretch, $\nu_3 = 2349$ cm$^{-1}$.  Only the bending mode has a large direct effect on global warming of Earth, but the other two modes have indirect effects since they have ``Fermi resonances" with overtones of the bending mode.}
\label{Fig1}
\end{center}
\end{figure}

The situation with polyatomic molecules is more complicated and interesting. To illustrate the basic ideas, we will discuss the triatomic CO$_2$ molecule in some detail.  Fig. \ref{Fig1} shows the main features of the CO$_2$ molecule, as they were sketched by Enrico Fermi in his classic paper on mode mixing and “Fermi resonances,” published in the year 1931 [5].   Similar considerations apply to other polyatomic molecules like methane or nitrous oxide. The unexcited CO$_2$ molecule, with neither rotational nor vibrational energy, is linear.   The C atom in the center has a slight positive charge because of the strongly electronegative O atoms on each end. The CO$_2$ molecule has no electric dipole moment because the centers of positive and negative charge both coincide with the center of symmetry. 

A CO$_2$ molecule can bend and vibrate after being hit by another molecule during a collision, much like a xylophone bar will vibrate after being hammered by a musician. Just as the vibrating xylophone bar emits a sound wave into the air, the vibrating CO$_2$ molecule emits thermal radiation. A bent CO$_2$ molecule has an electric dipole moment that points from the center of mass of the two, slightly negative O atoms toward the slightly positive C atom. The vibrations of this bending mode make the biggest contribution to the absorption and emission of thermal radiation by CO$_2$ molecules.  

As sketched in Fig. \ref{Fig1}, the CO$_2$ molecule also has two modes of vibration that keep the straight, linear alignment of the three atoms in the unexcited molecule. In the symmetric stretch mode, the C atom remains fixed as the O atoms vibrate with equal and opposite displacements along the molecular axis. For this mode of vibration, there is no vibrating dipole moment, since the centers of positive and negative charge remain at the center of the molecule. So the symmetric stretch mode of CO$_2$ does not contribute directly to the greenhouse properties of the molecule, although it can have an indirect influence through “mode mixing.” For the asymmetric stretch mode of CO$_2$, the carbon atom moves in one direction along the symmetry axis, and the two O atoms on either side move in the opposite direction. The vibration frequency of the asymmetric stretch mode is higher than the frequencies of most thermal photons, so it has little direct effect on the greenhouse properties of the molecule. However, the resulting strong absorption band, centered at a wavelength of 4.3 micrometers, is used for most non-dispersive infrared monitors that measure CO$_2$ concentrations in ambient air.

Molecules in Earth’s atmosphere not only vibrate but rotate.  This adds rotational sidebands to the vibration-frequencies of molecules. Because of the transverse nature of electromagnetic waves, vibrating molecules do not easily emit nor absorb radiation propagating along the direction of the vibrational axis. The vibrating electric dipole produces an anisotropic radiation pattern with most radiation emitted at right angles to the axis of vibration.  Rotating molecules produce a rotating radiation pattern, much like the light beam of a rotating searchlight or the radar beam from a rotating airport antenna. This causes the observed intensity to blink on and off at the rotation frequency. As in amplitude modulation of “carrier” frequencies of radio transmitters, upper and lower sidebands, displaced by the rotational frequency, are added to the vibrational carrier frequency of the radiation. The low-frequency sideband is called the P branch and the high-frequency side band is called the R branch. 

Electronic or nuclear excitations are needed to permit linear molecules like N$_2$, O$_2$, CO or CO$_2$ to rotate around their axial symmetry axis. The energies are much too high to be thermally excited in Earth’s atmosphere. So the rotation axis for linear molecules is always perpendicular to the symmetry axis. This causes the radiation patterns from axial vibrations of CO or CO$_2$ to split into the lower and upper sidebands that make up the P and R branches. There is no radiation at the vibration frequency of the non-rotating molecule. A CO$_2$ molecule vibrating in the asymmetric stretch mode radiates in much the same way as a “double-sideband carrier-suppressed” radio transmitter, sometimes used by ham radio operators.

Rotations of CO$_2$ molecules with bending-mode vibrations are more interesting. The rotation axis must still be perpendicular to the molecular symmetry axis.  But the molecule is vibrating at right angles to the symmetry axis, so the vibration axis may be perpendicular or parallel to the rotation axis. If the rotation and vibration axes happen to be perpendicular, as is always the case for asymmetric stretch vibrations, P and R sidebands are produced by the spinning antenna pattern. But if the vibrations happen to be parallel to the rotation axis, the radiation pattern will be unaffected by rotations, and the molecule will emit radiation at the unshifted vibrational frequency.   This is called the Q branch of the bending-mode band, and it is at the frequency of the bending mode. Slightly displaced sidebands are produced by centrifugal stretching of the rotating molecule, but the strongly displaced P and R branches are missing.   The Q branch of the bending mode of CO$_2$ can be seen clearly in high-resolution, spectral measurements of thermal radiation from Earth’s atmosphere.

The vibration-rotation spectrum of nitrous oxide, N$_2$O, another linear molecule like CO$_2$ has properties similar to those discussed above for CO$_2$.  But the analog of CO$_2$’s symmetric stretch mode for the less symmetric molecule N$_2$O can emit and absorb radiation, unlike CO$_2$ where this mode is “infrared inactive”.  The tetrahedral molecule methane, CH$_4$, is somewhat more complicated but like CO$_2$ it has several vibrational modes, some of which can absorb and emit radiation, and some of which cannot. Like the CO$_2$ bending mode, the methane bands have P, Q and R branches.

\begin{figure}[t]
\begin{center}
\includegraphics[height=100mm,width=.8\columnwidth]{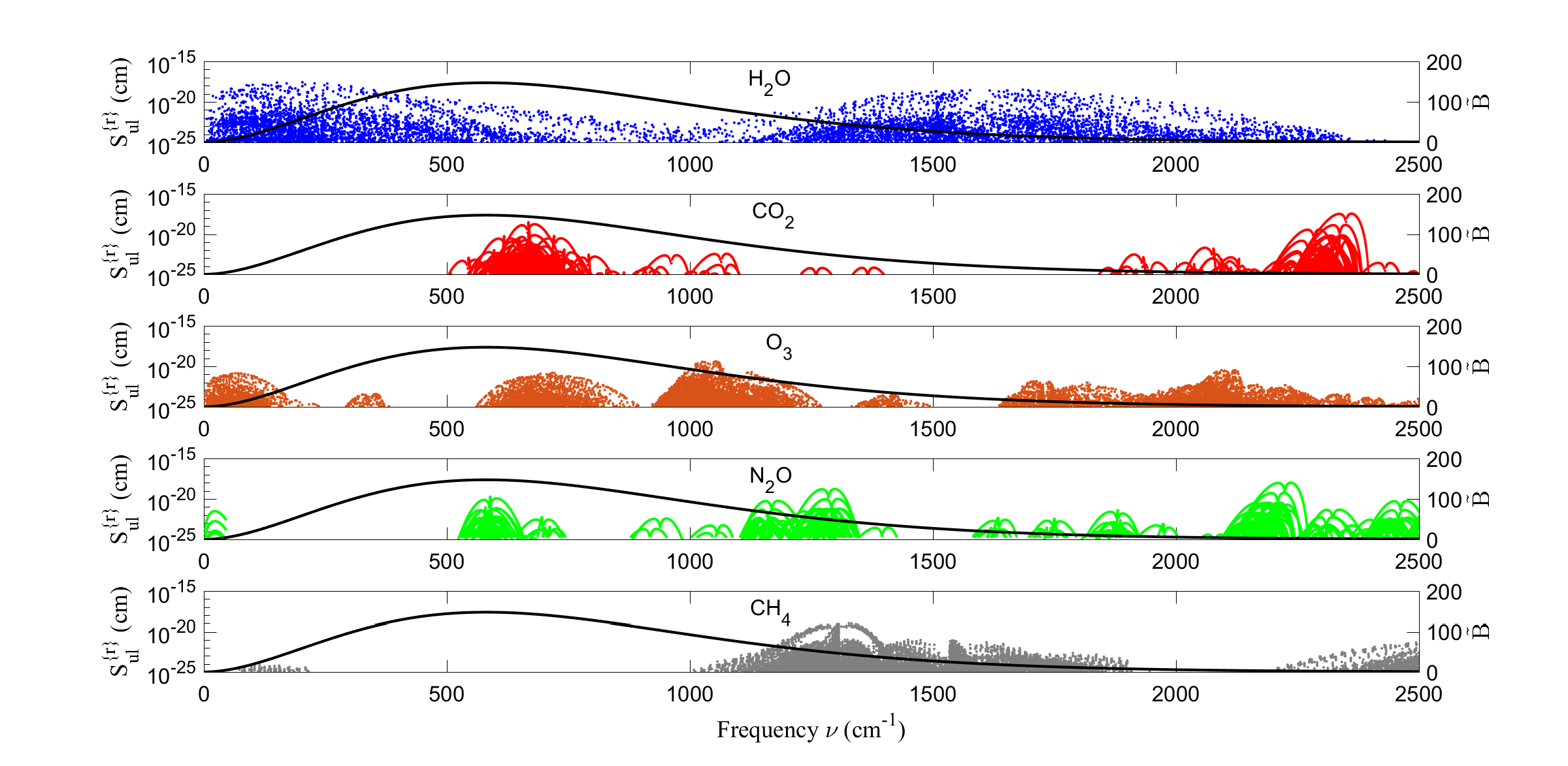}
\caption{To calculate greenhouse effects in detail, one must include the opacity of hundreds of thousands of individual line intensities, shown here as colored dots from the HITRAN data base.  More details of how line intensities are used can be found in references \cite{WvW1,WvW2}.}
\label{Fig2}
\end{center}
\end{figure}

\subsection{A Model Atmosphere}

How greenhouse gases affect Earth’s climate is a complicated issue, where atmospheric thermodynamics and convection are intimately involved. We will simplify the discussion as much as possible, but we will also try to adhere to Einstein’s admonition:  “Everything should be made as simple as possible, but not simpler”. We hope our readers will take the trouble to understand the basic thermodynamics that we review below.

Solar heating drives Earth’s climate. At the mean distance of Earth from the Sun, sunlight carries an energy flux of about 1,360 Watts per square meter (W/m$^2$). We are familiar with this flux, part of which warms us when we sunbathe at the beach on a cloud-free summer day. The flux at the top of Earth's atmosphere actually varies a little bit over the year, since Earth's orbit around the Sun is slightly elliptical. Earth is about 3.3\% closer to the Sun in early January than in early July. Since solar flux decreases as the square of the distance from the Sun, the solar flux at the top of the atmosphere is about 6.7\% or 91 W/m$^2$ greater in January than in July. As we will discuss in more detail below, for cloud-free temperate latitudes, doubling the concentration of CO$_2$ would decrease thermal radiation to space by about 3 W/m$^2$. Note 3 W/m$^2$ is much less than the planet-wide, winter-summer difference of 91 W/m$^2$.  And we are a long way from doubling CO$_2$ .

Launching into our maximally simplified discussion of the greenhouse effect, we consider a hypothetical Earth with a transparent atmosphere that is 80\% nitrogen and 20\% oxygen, and with the same mass as today’s atmosphere. But we assume no greenhouse gases at all, no CO$_2$, no H$_2$O and no clouds. To be consistent with no clouds, this hypothetical Earth must have no oceans, from which water vapor could evaporate. Oxygen, O$_2$, actually does absorb a very small amount of sunlight and also thermal radiation, but we will ignore that absorption and assume the atmosphere is completely transparent. To further simplify the problem, we assume that the Sun shines steadily with equal intensity on every part of Earth’s surface, from the tropics to the poles. Earth’s rotation provides daily averaging of insolation over longitude but not latitude. We will neglect the Coriolis force which is caused by the Earth's rotation, that is so important for dynamics of the real atmosphere.

With its radius, $r = 6378$ km, the Earth intercepts the solar flux with an equivalent disk area of $\pi r^2$. But the area of the spherical surface of the Earth is $4\pi r^2$.  So the average flux per unit area of Earth’s surface is one quarter of that at the top of the atmosphere or

\begin{equation}
F = 1360/4 {\rm\ W/m}^2  = 340 {\rm\ W/m}^2. 					
\label{eqn2}
\end{equation}

We assume the Earth is a perfect “blackbody” for both sunlight and thermal radiation, in the sense that the surface absorbs all radiation incident on it, irrespective of the frequency or direction of the incident light.  We assume negligible heat exchange between the surface and the interior of the Earth. Then heat can leave the surface in two ways:  

•	Heat can leave by thermal radiation through the transparent atmosphere to space;

•	Heat can flow to or from the atmosphere, either by conduction or convection, but not by radiation since we assume the atmosphere has no greenhouse gases and is completely transparent.

\subsection{An Isothermal Atmosphere}

We assume that the Earth has come to complete thermal equilibrium with an absolute surface temperature $T_0$, which is also equal to the temperature of the hypothetical, completely transparent atmosphere above.  The flux of thermal radiation by the Earth to space is equal to  the absorbed solar flux (\ref{eqn2}) so the surface neither heats nor cools. Then the Stefan-Boltzmann law of thermal radiation by blackbodies requires the thermal radiation emitted to space be

\begin{equation}
F = 340 {\rm\ W/m}^2  = \sigma T_0^4. 					
\label{eqn3}
\end{equation}

\noindent Here the Stefan Boltzmann constant is

\begin{equation}
\sigma= 5.67 \times 10^{-8} {\rm\ W/(m^2  K^4)}.			
\label{eqn4}
\end{equation}

\noindent The temperature that solves equation (\ref{eqn3}) is 

\begin{equation}
T_0 = 278.3 {\rm\ K}.						
\label{eqn5}
\end{equation}
	
Smaller temperatures for a greenhouse-free planet Earth are often quoted because clouds of condensed water reflect about 30\% of the solar radiation in an atmosphere paradoxically devoid of Earth’s most important greenhouse gas, water vapor. In discussions of the most basic physics of greenhouse warming, clouds are an unnecessary distraction. But clouds are very important for a detailed understanding of greenhouse effects.

Once the surface and the atmosphere have reached the same temperature, the temperature (\ref{eqn5}) of the surface will be the same value the Earth would have if there were no atmosphere at all. But the Earth’s atmosphere is massive, with a surface pressure of about 10 tons per square meter, the weight of air molecules in a one square meter column of air rising from the surface to infinity. For convenience, we assume the hypothetical atmosphere produces the “standard” surface pressure

\begin{equation}
p_0 = 10^5 {\rm\ Pa} = 1000 {\rm\ hPa}.					
\label{eqn6}
\end{equation}

\noindent The MKS unit of pressure, Pa = Pascale, is one Newton per square meter. Besides mass, the atmosphere also contains lots of energy, both thermal energy and gravitational potential energy. For computational convenience and physical clarity we will consider the energy per atmospheric molecule. The average thermal energy of a molecule of temperature $T_0$ is

\begin{equation}
u = c_v k_B T_0.
\label{eqn7}				
\end{equation}

\noindent Here the per-molecule heat capacity at constant volume, in units of Boltzmann’s constant (\ref{eqn1}), is very nearly 

\begin{equation}
c_v = 5/2.						
\label{eqn8}
\end{equation}

\noindent Recall from the kinetic theory of gases that the 5 in the numerator of (\ref{eqn8}) is the sum of 3, the number of translational degrees freedom, and 2, the number of rotational degrees of freedom for linear molecules like O$_2$ or N$_2$.  Atmospheric temperatures are not high enough for vibrational degrees of freedom to make a significant contribution to the heat capacity.  

For a static atmosphere, the air pressure at any altitude $z$ is simply the weight, per unit area, of air at higher altitudes. The differential expression of this fact is the equation of hydrostatic equilibrium,

\begin{equation}
dp = -\ {{mg}\over v} dz,
\label{eqn9}
\end{equation}

\noindent where the per-molecule gas volume is $v$ and $m$ is the mass per molecule. According to (\ref{eqn9}), the decrease $dp$ in pressure for an incremental increase $dz$ of altitude is the weight per unit area of an atmospheric slab of width $dz$.  Using the per-molecule ideal gas formula

\begin{equation}
pv = k_B T,
\label{eqn10}
\end{equation}

\noindent we can integrate (\ref{eqn9}) at constant temperature, $T=T_0$, to find the barometric formula for an isothermal atmosphere,

\begin{equation}
p = p_0 e^{-z/{z_T}}.
\label{eqn11}
\end{equation}

\noindent The average altitude of an atmospheric molecule in an isothermal atmosphere is

\begin{equation}
z_T = {{k_B T_0}\over {mg}}.
\label{eqn12}
\end{equation}

\noindent The mass per molecule of our model atmosphere of 80\% N$_2$ and 20\% O$_2$ by mole fraction is

\begin{equation}
m = 4.81 \times 10^{-26} {\rm \ kg}.
\label{eqn13}
\end{equation}

\noindent The average acceleration of gravity at Earth’s surface is

\begin{equation}
g = 9.8 {\rm \ m s^{-2}}.
\label{eqn14}
\end{equation}

\noindent So the average gravitational potential energy of a molecule in an isothermal atmosphere is

\begin{equation}
\epsilon_g = mg z_T = k_B T_0.
\label{eqn15}
\end{equation}

Summing the thermal energy (\ref{eqn7}) and the gravitational potential energy (\ref{eqn15}), we find the total, per-molecule energy of an isothermal atmosphere,

\begin{equation}
\epsilon = u + \epsilon_g = (c_v+1) k_B T_0 = c_p k_B T_0.		
\label{eqn16}
\end{equation}

\noindent Here the per-molecule heat capacity at constant pressure is 

\begin{equation}
c_p = c_v+1 = 7/2.
\label{eqn17}
\end{equation}

\noindent The per-molecule enthalpy is 

\begin{equation}
h = u + k_B T_0 = (c_v + 1)k_B T_0 = c_p k_B T_0,
\label{eqn18}
\end{equation}

\noindent the same as the formula (\ref{eqn16}) for the sum of the kinetic energy and gravitational potential energy. The identity of (\ref{eqn16}) and (\ref{eqn18}) is not accidental. The enthalpy increase $dH = N dh$ of a gas of $N$ molecules due to an increment $dQ$ of absorbed heat is the sum of the increase of thermal energy, $dU = N du$ and the work, $dW = p dV = N k_B dT$, done when the gas expands by a volume increment $dV = N dv$, at constant pressure $p$, into the surrounding environment.  A fraction $c_v/c_p  = 5/7$ of the heat added to an isothermal atmosphere goes into increasing the thermal energy of the molecules. A fraction $1/c_p= 2/7$ goes into lifting the molecules to higher altitudes. 

\subsection{Non-Isothermal Atmospheres}

For many reasons, the Earth’s atmosphere is not isothermal. Here we want to focus on the most important reason: greenhouse gases force Earth’s lower atmosphere, the troposphere, to convect. The convection drives the troposphere toward an adiabatic atmosphere, for which the dry adiabatic lapse rate is a temperature drop of $dT/dz = - 9.8$ K/km.  

To understand the essence of convection we use the concept of a fluid “parcel.” A parcel is a fixed mass $M$ of the fluid in a volume $V$. Noting that the number of air molecules in a mass $M$ is $N=M/m$, we can use the mass per molecule (\ref{eqn13}) and the ideal gas law (\ref{eqn10}) to show that a parcel with a mass $M = 1$ kg at the temperature (\ref{eqn5}) and pressure (\ref{eqn6}) would occupy a volume of $V = 0.78$ cubic meters (m$^3$). We assume that the parcel is small enough that all parts of it have nearly the same temperature $T$ and pressure $p$. 

The parcel is surrounded by other gas at nearly the same pressure $p$. If the parcel volume $V$ changes reversibly by the increment $dV$, the parcel will do an increment of work 

\begin{equation}
dW = p dV	
\label{eqn19}
\end{equation}

\noindent on the surroundings. The parcel can also reversibly exchange increments of heat, 

\begin{equation}
dQ = T dS,
\label{eqn20}
\end{equation}

\noindent with its surroundings. Heat is absorbed, $dQ > 0$ if the entropy increment of the parcel is positive, $dS > 0$, and heat is released, with $dQ < 0$ if the entropy change of the parcel is negative, $dS < 0$. 

The first law of thermodynamics is a statement of the conservation of energy. It can be written as

\begin{equation}
dU = dQ\ –\ dW = TdS - p dV,
\label{eqn21}
\end{equation}

\noindent The expression after the first equal sign is true in general, even for a non-reversible process where entropy increases inside the parcel with no addition of heat.  This often involves explosive processes like the ignition of the gas-air mixture of an internal combustion engine. The expression after the second equal sign is true for a reversible process like those of (\ref{eqn19}) and (\ref{eqn20}), where the “entropy of the universe” remains constant.

\subsection{Entropy}

Entropy is one of the most profound and poorly understood concepts of physics.  It is also very helpful in understanding how greenhouse gases work, so we briefly review it here. The most fundamental definition of the thermodynamic temperature follows from (20) and is, 

\begin{equation}
{1 \over T} = {{dS}\over {dQ}}.	
\label{eqn22}
\end{equation}

\noindent The ratio of the small entropy increase $dS$ of a parcel in thermal equilibrium to the small quantity of heat, $dQ$, that was absorbed to cause the entropy increase, is the inverse of the thermodynamic temperature.  Adding heat to a system can increase its temperature, so the heat increment $dQ$ must be small enough to cause a negligible increase of the temperature $T$.  In (\ref{eqn22}) we have assumed that no water contributes to the energy change of the parcel, $dW = 0$.

Like energy, entropy is additive, so we can compute the total entropy of the atmosphere by adding up the entropies of all of its parcels.  But unlike energy, which is rigorously conserved, “the entropy of the universe” is a steadily increasing quantity.  This is the essence of the Second Law of Thermodynamics.  The German physicist Clausius, who is credited, along with the Scottish physicist William Thomson (Lord Kelvin), with discovering the significance of entropy, summarized all of thermodynamics in the laconic words:
{\it Die Energie der Welt ist konstant. Die Entropie der Welt strebt einem Maximum zu.}
The energy of the universe is constant. Its entropy tends to a maximum \cite{Clausius}.

For example, suppose two air parcels are in contact, the first with absolute temperature $T_1$ and the second with a different temperature $T_2$. Transferring a small, positive quantity $dQ$ of heat from parcel 1 to parcel 2 will change the entropy of parcel 2 by $dS_2 = dQ/T_2$ and will change the entropy of parcel 1 by $dS_1 = -dQ/T_1$.  Since energy is conserved, the increment of heat $dQ$ gained by parcel 2 is equal to the increment of heat lost by parcel 1. As a result of this exchange, the entropy of the universe changes by 

\begin{equation}
dS = dS_1 + dS_2 = dQ \bigg({{-1}\over {T_1}} +{1\over {T_2}}\bigg).
\label{eqn23}
\end{equation}

\noindent We have assumed a positive heat increment, $dQ > 0$, so to ensure that $dS > 0$, (\ref{eqn22}) implies that $1/T_2 > 1/T_1$; that is, $T_1 > T_2$.  Heat can flow spontaneously from a hot parcel to a cold parcel, but never in the opposite direction.  An exception is exceptionally small parcels where “fluctuations” allow heat to flow temporarily from cool surroundings into a warmer, small parcel. An extreme example is a “parcel” consisting of a single CO$_2$ molecule which regularly gets so much hotter than the surrounding gas that it can emit a photon at the frequency of a vibrational mode, even though the photon energy is several times larger than $k_B T$.

The concept of entropy originated in classical thermodynamics, in the mid-19th century. But it played a key role in Boltzmann’s development of classical statistical physics, and in Planck’s invention of quantum mechanics toward the end of the century.  It was Boltzmann who first recognized the close connection between entropy $S$ and the thermodynamic probability, $W$. Here $W$ stands for Wahrscheinlichkeit = probability in German, not for work.  Boltzmann considered the connection so important, and rightly so, that engraved on his tombstone in Vienna’s Zentralfriedhof is the equation

\begin{equation}
S = k{\rm \ } log W.
\label{eqn24}
\end{equation}

\noindent Here $k = k_B$ is Boltzmann’s constant of (\ref{eqn1}), the natural unit of entropy. 

In statistical mechanics, entropy is a measure of disorder of the physical system. For example, one kg of boiling water at a temperature of 100 C, a pressure of 1 atmosphere and volume of 1 liter, has far less entropy or disorder than one kg of steam, which has the much larger volume of 1244 liters at the same temperature and pressure.

Entropy is an extensive state function of thermodynamic systems, much like the volume $V$ or the internal energy $U$.  The units of entropy are J/K, Joules per degree Kelvin, the same units as Boltzmann’s constant (\ref{eqn1}).  For an air parcel with $N$ molecules and total entropy $S$, we will define the entropy per molecule as

\begin{equation}
s = {S \over N}.	
\label{eqn25}
\end{equation}

\noindent Then the first law of thermodynamics (\ref{eqn21}) for an ideal gas can be written in the per-molecule form

\begin{equation}
c_v k_B dT = T ds\ –\ p dv.					 	
\label{eqn26}
\end{equation}

\noindent Integrating (\ref{eqn26}) with the aid of the per-molecule ideal gas law (\ref{eqn10}), we find the per-molecule entropy

\begin{equation}
s = s_o + c_p  k_B ln(T/T_0)\ –\ k_B ln (p/p_0).		  	    					 	
\label{eqn27}
\end{equation}

\noindent The per-molecule entropy at the Earth’s surface is $s_o$. The per-molecule heat capacity at constant pressure $c_p$ was given by (\ref{eqn17}). The pressure and temperature at the surface are $p_0$ and $T_0$. 

For an isothermal atmosphere, $T = T_0$ at all altitudes $z$, the pressure is given by the barometric formula (\ref{eqn11}).  So the per-molecule entropy (\ref{eqn27}) for an isothermal atmosphere increases linearly with altitude and is given by

\begin{equation}
s = s_o + {{mgz}\over {T_0}}.		  	    					 	
	\label{eqn28}
\end{equation}

For a general, non-isothermal atmosphere we can differentiate (\ref{eqn27}) with respect to altitude $z$, and use the equation of hydrostatic equilibrium (\ref{eqn9}) for $dp/dz$, to find

\begin{equation}
{{ds}\over {dz}} = {{c_p k_B}\over T} \bigg({{dT}\over {dz}} + L\bigg),		  	    					 	
\label{eqn29}
\end{equation}

\noindent In (\ref{eqn29}) we have introduced the “adiabatic lapse rate” for dry air 

\begin{equation}
L = {{mg}\over{c_p k_B}} = 9.8 {\rm\ K/km}.				 	
\label{eqn30}
\end{equation}

In atmospheric physics the entropy per molecule is often defined in terms of a function of temperature and pressure called the “potential temperature,”

\begin{equation}
\theta = T \bigg({{p_0}\over  p}\bigg)^{1/c_p}.
\label{eqn31}
\end{equation}

\noindent In terms of the potential temperature (\ref{eqn31}) the expression (\ref{eqn27}) of the per-molecule entropy simplifies to

\begin{equation}
s = s_o + k_B ln \bigg({{\theta}\over {T_0}}\bigg)^{c_p}.	
\label{eqn32}
\end{equation}

\noindent Molecules with the same potential temperature  have the same entropy $s$.

\subsection{Atmospheric Stability}

The linear increase of entropy with altitude of an isothermal atmosphere, which is described by (\ref{eqn28}) gives the atmosphere “stability,” in the sense that air parcels try to remain at their original altitudes. To get a more quantitative understanding of atmospheric stability we note that the mass density of environmental air is

\begin{equation}
\rho = {m\over v}  = {{mp}\over {k_B T}}.	
\label{eqn33}
\end{equation}

\noindent According to the Archimedes Principle, the per-molecule
force $f$ on a parcel p of mass density $\rho_{\rm p} = m/v_{\rm p}$ 
and per-molecule volume $v_{\rm p}$, that has displaced an equal
volume $v = v_{\rm p}$ of environmental air of mass density $\rho$ is

\begin{equation}
f = - gv_{\rm p} (\rho_{\rm p} - \rho).	
\label{eqn34}
\end{equation}

\noindent The first term of (\ref{eqn34}), $- g v_{\rm p} \rho_{\rm p} = - gm$, is the negative force of gravity per parcel molecule. 
The second term, $gv_{\rm p} \rho$, is the upward buoyant force of displaced environmental air.  The buoyant force will not be 
big enough to support the parcel’s weight if $\rho < \rho_{\rm p}$ and the parcel will sink down.  Or the buoyant force may exceed the parcel’s weight if $\rho > \rho_{\rm p}$ and the parcel will float up. Suppose that at altitude $z$, the parcel is a sample volume of environmental air, so that $\rho_{\rm p}(z) = \rho(z)$. Then from (\ref{eqn34}), $f(z) = 0$ and the parcel will be in mechanical equilibrium with no net force acting on it. 

Now imagine lifting a parcel of environmental air by an altitude increment $dz$, where the environmental air density is $\rho(z+dz) = \rho(z) + d\rho$. Since the parcel is neutrally buoyant, negligible work is needed to lift it by a small amount. The lifted parcel will expand and do an increment $dW$ of work on the lower-pressure surrounding air. From the first law of thermodynamics the energy for the work comes from heat $dQ$ that flows into the parcel, and from $-dU$, the decrease  of internal energy. Because of the very small thermal conductivity of air, it takes a very long time for appreciable heat to flow into or out of a parcel of reasonable size. For example, temperature equilibration by molecular heat conductivity alone in a parcel with a diameter $d = 1$ meter would require a time of order, $t = d^2 / D$, or about a day for a sea-level molecular diffusion coefficient of $D = 0.1$ cm$^2$ s$^{-1}$.  But the pressure equilibration times are on the order of $d / v$, the parcel diameter divided $d$ by the speed of sound, $v$, around 300 m/s. The pressure equilibration time for the parcel would only be a few milliseconds.  We will therefore consider lifting processes where there is not enough time for appreciable heat flow into or out of a parcel, and we can set $dQ = 0$. Then the lifted parcel retains the same per-molecule entropy as it had at the altitude $z$, and the lifting will be adiabatic.

At the altitude $z+dz$ the force may no longer be zero since the lifted parcel may have a different density from the environmental air it displaced. The net per-molecule force at the altitude $z+dz$ will be

\begin{equation}
df = -gv (d\rho_{\rm p} - d\rho).
\label{eqn35}
\end{equation}

\noindent The density increment of the adiabatically lifted air is $d\rho_{\rm p}$ and the density increment of the displaced air is $d\rho$. Differentiating the expression (\ref{eqn33}) and using the condition (\ref{eqn9}) for hydrostatic equilibrium as well as the ideal gas law (\ref{eqn10}), we find that the density increment for the environmental air is

\begin{equation}
d\rho = - {m \over{Tv}} \bigg({{mg}\over {k_B}} + {{dT}\over {dz}}\bigg) dz.
\label{eqn36}
\end{equation}

\noindent To order $dz$ we get the density increment for the adiabatically lifted parcel from (\ref{eqn36}) by replacing $dT/dz$ by the negative of the adiabatic lapse rate (\ref{eqn30}),

\begin{equation}
d\rho_{\rm p} =- {m \over {Tv}} \bigg({{mg}\over {k_B}} - L\bigg) dz.	
\label{eqn37}
\end{equation}

\noindent Substituting (\ref{eqn36}) and (\ref{eqn37}) into (\ref{eqn35}), as well as using (\ref{eqn29}) and ({\ref{eqn30}), we find the force per parcel molecule

\begin{equation}
df = -{{mg} \over T} \bigg(L + {{dT}\over {dz}}\bigg) dz = -{{mg} \over {c_p k_B}}\ {{ds} \over {dz}}\ dz=-L ds.		
\label{eqn38}
\end{equation}

\noindent If a parcel is adiabatically lifted by an increment $dz$ in altitude, the increment $df$ of the buoyant force per molecule is simply the negative product of the adiabatic lapse rate $L$ of (\ref{eqn30}) and the per-molecule entropy increment, $ds$, of environmental air between altitudes $z + dz$ and $z$.  If the per-molecule entropy increases with increasing altitude, $ds/dz > 0$, there will be a net restoring force that will try to push the parcel back down to its original altitude, or push it back up if $dz < 0$.	

For a stable atmosphere, with $ds/dz > 0$, the molecules behave as if they were bound to their initial altitudes by springs of force constants $k = L ds/dz$.  We recall from
elementary mechanics that a particle of mass $m$ held by a spring of force constant $k$ will oscillate with a frequency
$\Omega = (k/m)^{1/2}$.  So a stable atmosphere has a buoyancy frequency $\Omega$ given by

\begin{equation}
\Omega^2 = {L \over m}\ {{ds}\over {dz}} = {g \over {c_p k_B}}\ {{ds}\over {dz}}.		
\label{eqn39}
\end{equation}

\noindent The buoyancy frequency, $\Omega$ defined by (\ref{eqn39}), is called the Brunt-Vaisala frequency \cite{Brunt}.

For an isothermal atmosphere, with $dT/dz = 0$, we can use the expression (\ref{eqn29}) for $ds/dz$ to write the square of the buoyant frequency as   

\begin{equation}
\Omega^2 = {{mg^2}\over {c_p k_B T_0}} = {{g L}\over {T_0}}.	
\label{eqn40}
\end{equation}

\noindent Using the acceleration of gravity $g$ of (\ref{eqn14}), the adiabatic lapse rate $L$ and the surface temperature $T_0$ of (\ref{eqn5}) in (\ref{eqn40}), we find that the buoyancy frequency of a hypothetical, isothermal atmosphere is

\begin{equation}
\Omega = 0.0186 {\rm\ s}^{-1} = {{2\pi} \over{ 338 {\rm\ s}}}		
\label{eqn41}
\end{equation}

\noindent The oscillation period is 338 s, or between 5 and 6 minutes. This period is comparable to the fluctuation times of the internal structure of clouds, drifting through a blue sky.
We recall that the squared frequency of a pendulum of length $l$ is 

\begin{equation}
\Omega^2 = {g \over l}.
\label{eqn42}
\end{equation}

\noindent Comparing (\ref{eqn40}) and (\ref{eqn42}), we see that the buoyancy oscillations of an isothermal atmosphere have the same frequency as a pendulum of length 

\begin{equation}
l = {{T_0}\over L} = z_a = 28.4 {\rm\ km}.					
\label{eqn43}
\end{equation}

\noindent As we will discuss in the following section, the length $z_a$ defined by (\ref{eqn43}) is the height of an adiabatic  atmosphere of ideal gas, with a surface temperature $T_0$. The numerical value, 28.4 km, would be in the mid stratosphere of Earth’s atmosphere.

\subsection{An Adiabatic Atmosphere}

In addition to the isothermal atmosphere discussed above, another important limit is the adiabatic atmosphere, where the per-molecule entropy $s$ is independent of altitude $z$. For parcels in Earth’s atmosphere, we will use the adjectives adiabatic and isentropic interchangeably, but under some conditions, adiabatic processes can differ dramatically from isentropic processes.  An isentropic change in the volume of a parcel is one where the parcel entropy is the same before and after the change. An adiabatic change of the volume of a parcel is one in which no heat is exchanged between the parcel and the environment. But the moving boundaries of the parcel can do positive or negative work on the parcel.   

Adiabatic volume changes cannot be too fast or they will increase the parcel entropy.  For example, if the containing boundaries move at supersonic speeds, converging boundaries can produce shock waves, and expanding boundaries can produce expansion fans in the parcel gas. Both phenomena increase the parcel entropy, even though no heat has been added.  This is an important issue for laser-fusion work, where the parcel is the gas in a small pellet that is imploded by intense laser pulses. Most natural atmospheric changes are slow enough that isentropic and adiabatic mean the same thing.

From inspection of (\ref{eqn29}) we see that for an adiabatic atmosphere, with $ds/dz = 0$, the rate of change of temperature with altitude must be equal and opposite to the adiabatic lapse rate,  

\begin{equation}
{{dT}\over{dz}} = - L = - 9.8 {\rm\ K/km}.					
\label{eqn44}
\end{equation}

\noindent The temperature profile of an adiabatic atmosphere must therefore be

\begin{equation}
T = T_a - z L.					
\label{eqn45}
\end{equation}

\noindent The temperature, $T_a$, at the bottom of the adiabatic atmosphere need not be the same as the equilibrium surface temperature $T_0$ of (\ref{eqn5}) for an atmosphere without greenhouse gases.  The thickness, $z_a$, of an adiabatic atmosphere,

\begin{equation}
z_a= {{T_a}\over L} = {{c_p k_B T_a}\over {mg}}				
\label{eqn46}
\end{equation}

\noindent is the altitude at which the absolute temperature (\ref{eqn45}) goes to zero, $T = 0$. We already mentioned in connection with (\ref{eqn43}) that a representative thickness is $z_a= 28.4$ km.

Using the adiabatic temperature profile (\ref{eqn45}) with the formula (\ref{eqn27}) for the per-molecule entropy and replacing the surface temperature $T_0$ with $T_a$, we find that the pressure in an isentropic atmosphere must be given by the formula

\begin{equation}
p = p_0 (1-z/z_a)^{c_p}.							
\label{eqn47}
\end{equation}

\noindent The fraction of atmospheric molecules between altitudes with pressure $p$, and altitudes with pressure $p+dp$, is $dp/p_0$. So the average temperature of a molecule in an adiabatic atmosphere is

\begin{equation}
< T > =  {1 \over {p_0}} \int_0^{p_0} T dp.					
\label{eqn48}
\end{equation}

\noindent Noting from (\ref{eqn45}) and (\ref{eqn47}) that

\begin{equation}
T=T_a \bigg({p\over{p_0}}\bigg)^{1/c_p},				
\label{eqn49}
\end{equation}

\noindent we can integrate (\ref{eqn48}) to find 

\begin{equation}
< T > =T_a {{c_p}\over{c_p+1}}.
\label{eqn50}
\end{equation}

\noindent We can replace $T_0$ with $< T >$ in (\ref{eqn7}) to find that the per-molecule thermal energy of an adiabatic atmosphere is

\begin{equation}
u =c_v k_B T_a {{c_p} \over{c_p+1}}.
\label{eqn51}
\end{equation}

\noindent In like manner, one can show that the mean altitude of a molecule is

\begin{equation}
<z> = {1\over{p_0}}  \int_0^{p_0}z dp = {{z_a}\over{c_p+1}}.
\label{eqn51b}
\end{equation}

\noindent Then the mean per-molecule gravitational energy becomes

\begin{equation}
\epsilon_g = mg<z>= k_B T_a {{c_p}\over {c_p+1}}.
\label{eqn52}
\end{equation}

For an adiabatic atmosphere with a surface temperature $T_a$ equal to the surface temperature $T_0$ of an isothermal atmosphere, both the thermal and gravitational energies of the adiabatic atmosphere are smaller than those of the isothermal atmosphere by the factor $c_p/(c_p+1)$ or 7/9 for the hypothetical atmosphere of diatomic molecules with negligible vibrational excitation. For the same surface temperature, molecules of an adiabatic atmosphere have a lower average altitude and a lower average temperature than those of an isothermal atmosphere.

Finally, we note that since $ds/dz = 0$ for an adiabatic atmosphere, the buoyancy frequency for the atmosphere, given by (\ref{eqn39}), is zero,

\begin{equation}
\Omega = 0.
\label{eqn53}
\end{equation}

\noindent Air parcels in an adiabatic atmosphere have no tendency to return to their original altitude if they are displaced upward or downward. There is no restoring force, as there is for an isothermal atmosphere.

In the absence of greenhouse gases, the isothermal atmosphere will not change with time, since there is no thermal gradient to drive heat flow. However, without greenhouse gases, a thermally isolated adiabatic atmosphere would slowly evolve to an isothermal atmosphere because of conductive heat flow from the warmer lower atmosphere to the colder upper atmosphere.  

As we will discuss in the next section, atmospheres with sufficient concentrations of greenhouse gases can maintain an approximately adiabatic troposphere and an approximately isothermal stratosphere indefinitely, because greenhouse gases facilitate convective transfer of heat from the solar heated surface through the troposphere.

\subsection{Atmospheres with Greenhouse Gases}

The physics we have outlined above changes dramatically if we add greenhouse gases to our model. For simplicity, we will consider a well-mixed greenhouse gas, like CO$_2$ in Earth’s atmosphere.  In keeping with our earlier discussion of greenhouse gases, our model atmosphere with greenhouse gas lets the Sun continue to heat the Earth’s surface with the flux (\ref{eqn3}), but the greenhouse gases attenuate thermal radiation such that the fraction of monochromatic vertical radiation that reaches outer space without absorption is

\begin{equation}
f = e^{-\tau_o}.
\label{eqn54}
\end{equation}

\noindent Here $\tau_o$, the vertical optical depth between the surface and outer space, will be proportional to the partial pressure of the greenhouse gases.  For Earth and other planets of the Sun, the optical depth can greatly exceed unity at the centers of strong absorption bands. Negligible surface radiation reaches outer space for these strongly absorbed frequencies, and most escaping thermal radiation is emitted by greenhouse gases at altitudes well above the surface.  

The radiation to space by greenhouse gases cools air parcels. The cooling is most pronounced at higher altitudes where the radiation has a good chance to reach outer space without being absorbed by still higher-altitude greenhouse gases. Therefore, greenhouse gases cause high-altitude air parcels to cool. As the parcels cool, contract and become heavier than the surrounding air, they sink and are replaced by rising parcels of warmer air that float up from below. 

Since greenhouse gases do not absorb solar radiation, the Sun continues to heat the surface at the rate (\ref{eqn3}). The rising and sinking air parcels exchange little heat with the surrounding air, so they behave like an adiabatic conveyor belt that carries the solar energy that has been absorbed by the surface to sufficiently high altitudes where radiating greenhouse gases can release the energy to space. The most dramatic effect of sufficiently high concentrations of greenhouse gases is to create a nearly adiabatic troposphere of convecting air. 

Because of the complicated vibrations and rotations of real greenhouse molecules, the optical depth $\tau_o$ of (\ref{eqn54}) depends strongly on the frequency of the thermal radiation. There are infrared “window” frequencies where there is almost no attenuation and one can set $\tau_o = 0$.   But for frequencies in the center of the Q branch of CO$_2$, one finds optical depths of order $\tau_o = 500,000$.

\section{A Gray Atmosphere}

A model for the frequency dependence of atmospheric opacity (which pushes the limits of Einstein’s admonition -- that models should not be simpler than possible) is a gray atmosphere, with no frequency dependence at all. Then the optical depth is the same for all thermal radiation frequencies.  We will also assume that as one descends into the atmosphere from outer space, where the pressure is $p = 0$, to an altitude of pressure $p$, the optical depth grows to the value

\begin{equation}
\tau=\tau_o {p \over{p_0}}.
\label{eqn55}
\end{equation}

\noindent The surface pressure $p_0$ was given by (\ref{eqn6}).  For an adiabatic atmosphere, we can use (\ref{eqn47}) to write the optical depth (\ref{eqn55}) as

\begin{equation}
\tau=\tau_o \bigg(1-{z \over{z_a}}\bigg)^{c_p}.
\label{eqn56}
\end{equation}

\noindent For optically thick atmospheres, with $\tau_o >> 1$, most of the monochromatic radiation from the atmosphere to space comes from altitudes $z$, or pressures $p$, near those for which the optical thickness to space is $\tau = 1$.  So we can use (\ref{eqn55}) to define emission pressure as

\begin{equation}
p_e = {{p_0}\over {\tau_o}}
\label{eqn57}
\end{equation}

\noindent and we can use (\ref{eqn56}) to define an emission altitude as

\begin{equation}
z_e = z_a \bigg(1 - {1\over {{{\tau_o}^{1/c_p}}}}\bigg).
\label{eqn58}
\end{equation}

At altitudes above (\ref{eqn58}) so little greenhouse gas remains that the radiative flux remains constant versus increasing altitude. The radiation flux below the emission altitude rapidly drops to nearly zero because the intensity becomes isotropic, with nearly as much downward radiative flux from the atmosphere above as upward flux from the atmosphere below. The emission level is a site of intense radiative cooling of the atmosphere. The greenhouse molecules at the emission altitude supply the model planet’s thermal radiation to space. Energy to replace that radiated away at the emission altitude is provided by convection of warm air parcels floating up from below. Having cooled, the parcels sink back to the surface to complete the convection cycle.

For the model of a gray atmosphere, the emission altitude (58) can be considered the tropopause, below which there is a convective troposphere, and above which there is a nearly isothermal stratosphere, with negligible convection. For this maximally simplified model, heat transport from the surface to the tropopause is almost all convective and heat transport above the troposphere is all radiative.

According to (\ref{eqn45}) and (\ref{eqn46}), the temperature at the emission altitude is

\begin{equation}
T_e = T_a\bigg(1 - {{z_e}\over{z_a}}\bigg).
\label{eqn59}
\end{equation}

\noindent Here $T_a$ is the surface temperature of an adiabatic atmosphere.  The emission altitude $z_e$ of (\ref{eqn58}) is where the gray adiabatic atmosphere sends the absorbed solar flux (\ref{eqn3}) back to space as thermal radiation. So the emission temperature $T_e$ must be the same as the surface temperature $T_0$ of (\ref{eqn5}) for an Earth with no greenhouse gases. So (\ref{eqn59}) implies that

\begin{equation}
T_0 = T_a\bigg(1 - {{z_e}\over{z_a}}\bigg).	
\label{eqn60}
\end{equation}

\noindent From (\ref{eqn60}) and (\ref{eqn58}) we find the surface temperature of the gray atmosphere is 

\begin{equation}
T_a = T_0 {\tau_o}^{1/c_p}.
\label{eqn61}
\end{equation}

\noindent Differentiating this equation yields

\begin{equation}
{{dT_a}\over {T_a}} = {{d\tau_o}\over {c_p \tau_o}}.
\label{eqn61b}
\end{equation}

\noindent All but 1\% of Earth's atmosphere is comprised of diatomic gases N$_2$ and O$_2$ for which $c_p =3.5$.  Hence, the relative change in surface temperature is 29\% of the relative change of the atmosphere's optical depth.  From (\ref{eqn46}), (\ref{eqn60}) and (\ref{eqn61}) we see that the emission (or tropopause) altitude of the gray atmosphere is

\begin{equation}
z_e = ({\tau_o}^{1/c_p} - 1) {{c_p k_B T_0}\over{mg}}.
\label{eqn62}
\end{equation}

The pressure at Earth’s tropopause varies from around 100 hPa in the tropics to 300 hPa at the poles. Taking a midlatitude pressure of $p_e = 200$ hPa with the surface pressure $p_0 = 1000$ hPa of (\ref{eqn6}), we see from (\ref{eqn57}) that the optical depth of our gray atmosphere would have to be 

\begin{equation}
\tau_o = 5.
\label{eqn63}
\end{equation}

\noindent According to (\ref{eqn61}), for the heat capacity $c_p = 3.5$ of (\ref{eqn17}) the optical depth (\ref{eqn63}) would require the surface to warm by a factor

\begin{equation}
{\tau_o}^{1/c_p} =  5^{1/3.5} = 1.58.	
\label{eqn64}
\end{equation}

\noindent For the surface temperature $T_0 = 278.3$ K of a model planet with no greenhouse gases, the temperature increase would then be

\begin{equation}
\Delta T = T_a - T_0 = ({\tau_o}^{1/c_p} - 1)T_0 = 162 {\rm\ K}.
\label{eqn65}
\end{equation}

\begin{figure}[t]
\begin{center}
\includegraphics[height=100mm,width=.8\columnwidth]{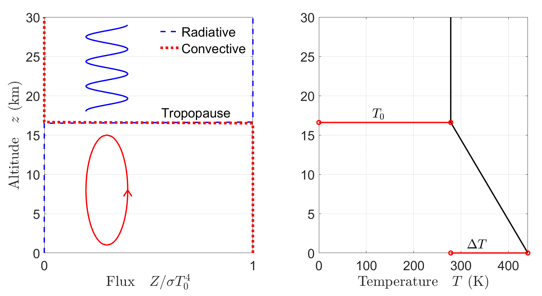}
\caption{The simplest example of greenhouse warming. Greenhouse gases in a gray atmosphere, which is uniform in the horizontal direction, produce a convecting, adiabatic troposphere below the tropopause, at the emission altitude $z_e$ of (\ref{eqn66}).  The emission temperature $T_0$ is the same as the surface temperature (\ref{eqn5}) of the surface if there were no greenhouse gases. In the adiabatic troposphere, the heat flux $Z$ is convective and is equal to the solar heating flux $F = \sigma T_0^4$ of (\ref{eqn3}). The heat flux in the troposphere is entirely convective while in the isothermal stratosphere the flux is entirely radiative.}
\label{Fig3}
\end{center}
\end{figure}

\noindent Using (\ref{eqn64}) in (\ref{eqn62}) we find a tropopause altitude for the gray atmosphere 

\begin{equation}
z_e = 16.6 {\rm\ km}.
\label{eqn66}
\end{equation}

\noindent The height of the troposphere in the real atmosphere of the Earth ranges from as low as 6 km over the midwinter poles to almost 20 km over the equator. 

The results of this section are summarized in Fig. \ref{Fig3}. The simple “one-dimensional” model of a gray atmosphere is only semi-quantitative, but it leads us to a number of important insights on how greenhouse gases work.

•	Greenhouse gases cool the upper atmosphere by radiating heat to space.

•	For a gray atmosphere, surface radiation to space is attenuated by factor $e^{-\tau_o}$ where $\tau_o$ is the vertical optical depth from the surface to outer space. Absorbed surface radiation is replaced by radiation emitted by greenhouse gases higher in the atmosphere.

•	For optical depths, $\tau_o >> 1$, atmospheres develop a lower troposphere where convection by parcels of warm air floating upward and parcels of cold air sinking downward transport most of the solar energy absorbed by the surface. Greenhouse gases radiate heat to space from altitudes close to the tropopause. Radiation to space from the surface and from greenhouse molecules in the lower troposphere is negligible.

•	Radiative heat transport is negligible compared to convective heat transport below the tropopause.

•	Convective heat transport is negligible compared to radiative heat transport above the tropopause.

•	Without greenhouse gases, the adiabatic temperature profile of the troposphere would evolve into an isothermal profile with the same temperature as the surface.

•	Greenhouse gases increase the temperature of the surface, compared to the surface temperature in an atmosphere without greenhouse gases.

\section{Realistic Model of Atmosphere}

Earth’s real atmosphere responds to greenhouse gases in a more complicated way than the simple model discussed above. But as the model predicts, the Earth’s atmosphere is partitioned into a convecting troposphere between the surface and the tropopause, and a stratosphere above the tropopause where there is very little vertical convection.  There is greenhouse warming of the real Earth’s surface, but much less than for the model discussed above. We briefly discuss some of the most important details of the atmosphere that must be accounted for to make a realistic model.

The Earth is not uniformly heated by the Sun, as was assumed in the model. The midday heating is greatest at the subsolar latitudes, where the Sun is directly overhead at noon. The subsolar latitude reaches its most northern bound, the Tropic of Cancer at 23.5$^o$ north around 21 June, the northern summer solstice. The most southern bound, the Tropic of Capricorn at 23.5$^o$ south, is attained on the southern summer solstice, about 21 December.   

\begin{figure}[t]
	\begin{center}
		\includegraphics[height=100mm,width=1\columnwidth]{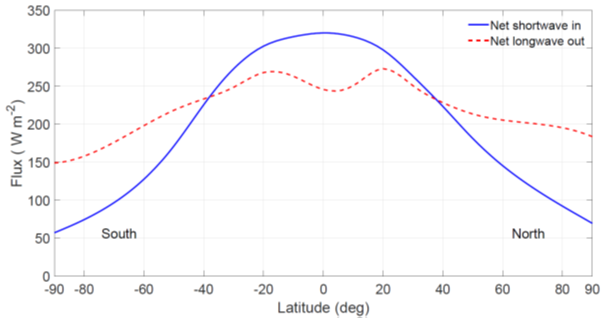}
		\caption{The continuous blue curve is the yearly average of incoming short-wave solar flux (net visible, near infrared and ultraviolet) absorbed by the Earth.  The dashed red curve is the yearly average of the outgoing thermal flux (net longwave infrared) radiated to space by the Earth.  Excess solar energy absorbed in the tropics is transported to the poles by mass flux in the atmosphere and oceans.  The data is from satellite observations \cite{Dewitte}.}
		\label{Fig4}
	\end{center}
\end{figure}

As shown in Fig. \ref{Fig4}, the yearly averaged solar insolation at the top of the atmosphere is about six times larger at the equator than at either pole.  A significant part of the solar energy absorbed in the tropics is convected by the atmosphere and oceans toward the poles before it is radiated to space by Earth’s surface and greenhouse gases.  In the tropics the Earth radiates less energy to space than the solar energy it receives, since lots of heat is exported north and south. Near the poles, the Earth radiates more heat to space than the solar energy it receives. Heat convected toward the poles from the tropics makes up the difference.

The intertropical convergence zone is where trade winds from the north and south converge. There, they loft huge quantities of moist air up to the stratosphere and produce heavy rainfall.  The cloud tops in the tropopause of the intertropical convergence zone can be up to 18 km high and with correspondingly low temperatures, down to 200 K or less. The weak thermal emission to space from the high, cold cloud tops of the tropics are responsible for the dip in the thermal flux at the top of the atmosphere, which can be seen in Fig. \ref{Fig4}. The intertropical convergence zone migrates north and south along with the subsolar latitude. This leads to the dramatic monsoons of tropical and subtropical latitudes.  The complicated distribution of land and oceans on Earth’s surface means that for a given longitude, the well-defined subsolar latitude may be greater or less than the latitude of the intertropical convergence zone.

Earth’s winds are strongly affected by the rotation of the Earth, which was also neglected in the simple model.  The Coriolis forces for the rotating Earth push moving currents of atmospheric air (or ocean water) to the right in the northern hemisphere and to the left in the southern hemisphere.  A dramatic result is the formation of jet streams in the upper atmosphere.

The preferential heating of the tropics causes atmospheric isobars, surfaces of constant pressure, to slope downward from the tropics to the poles.  The resulting “baroclinicity,” is a key driving factor for the Rossby waves of temperate latitude weather fronts and for tropical cyclones like hurricanes.

\begin{figure}[t]
\begin{center}
\includegraphics[height=100mm,width=.8\columnwidth]{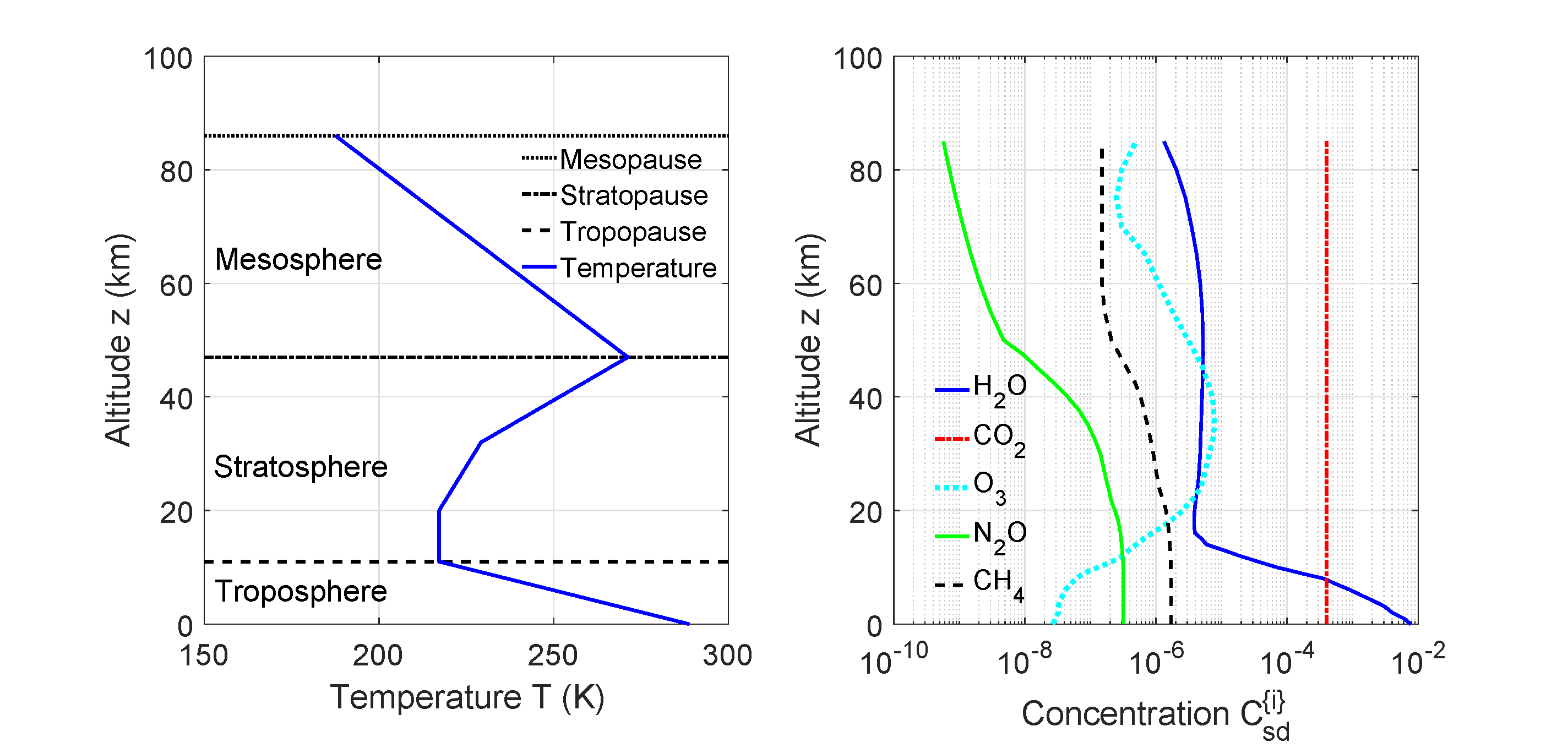}
\caption{{\bf Left.} A standard midlatitude atmospheric temperature profile \cite{USAtmos}.  The altitudes of the tropopause, stratopause and mesopause are shown.  Higher parts of the atmosphere have negligible effects on radiative transfer.  The Earth's mean surface temperature is 288.7 K.  {\bf Right.}  Observed concentrations, $C_{\rm sd}^{\lbrace i \rbrace}$ for greenhouse molecules versus altitude \cite{Anderson}.  The sea level concentrations are 7750 ppm of H$_2$O, 1.8 ppm of CH$_4$ and 0.32 ppm of N$_2$O.  The O$_3$ concentration peaks at 7.8 ppm at an altitude of 35 km.  The CO$_2$ concentration is 400 ppm at all altitudes.}
\label{Fig5}
\end{center}
\end{figure}

Unlike the model, which assumed dry air, the real atmosphere of the Earth is moist. The evaporation of water at the surface, the condensation of water to clouds in the troposphere, and the precipitation of rain and snow, transport large amounts of heat both vertically and horizontally. The condensation of moisture in upwardly convecting air parcels releases large amounts of heat to the surrounding N$_2$ and O$_2$ molecules that make up most of the atmosphere.  As a result, the dry adiabatic lapse rate of (\ref{eqn30}), $L = 9.8$ K/km, is only observed near the top of the troposphere, where little water vapor remains to condense. The lapse rate can be half the dry rate or less for very moist air near the surface. A typical average lapse rate between the surface and the tropopause is about 6.5 K/km, as shown in the “standard” temperature profile of the troposphere, shown in the left panel of Fig. \ref{Fig5}. 

Unlike the simple model, the per-molecule entropy, $s$, of the real troposphere increases slightly with increasing altitude, so the troposphere is not fully adiabatic and is marginally stable to vertical convection.  The model stratosphere is isothermal, but the temperature of the real stratosphere shown in Fig. \ref{Fig5} increases due to heating of the ozone layer by solar ultraviolet radiation. The rising temperature increases the restoring force $df$ of (\ref{eqn38}) for displaced air parcels. This increases the stability of the stratosphere to vertical convection. 

\begin{figure}[t]
	\begin{center}
		\includegraphics[height=100mm,width=1\columnwidth]{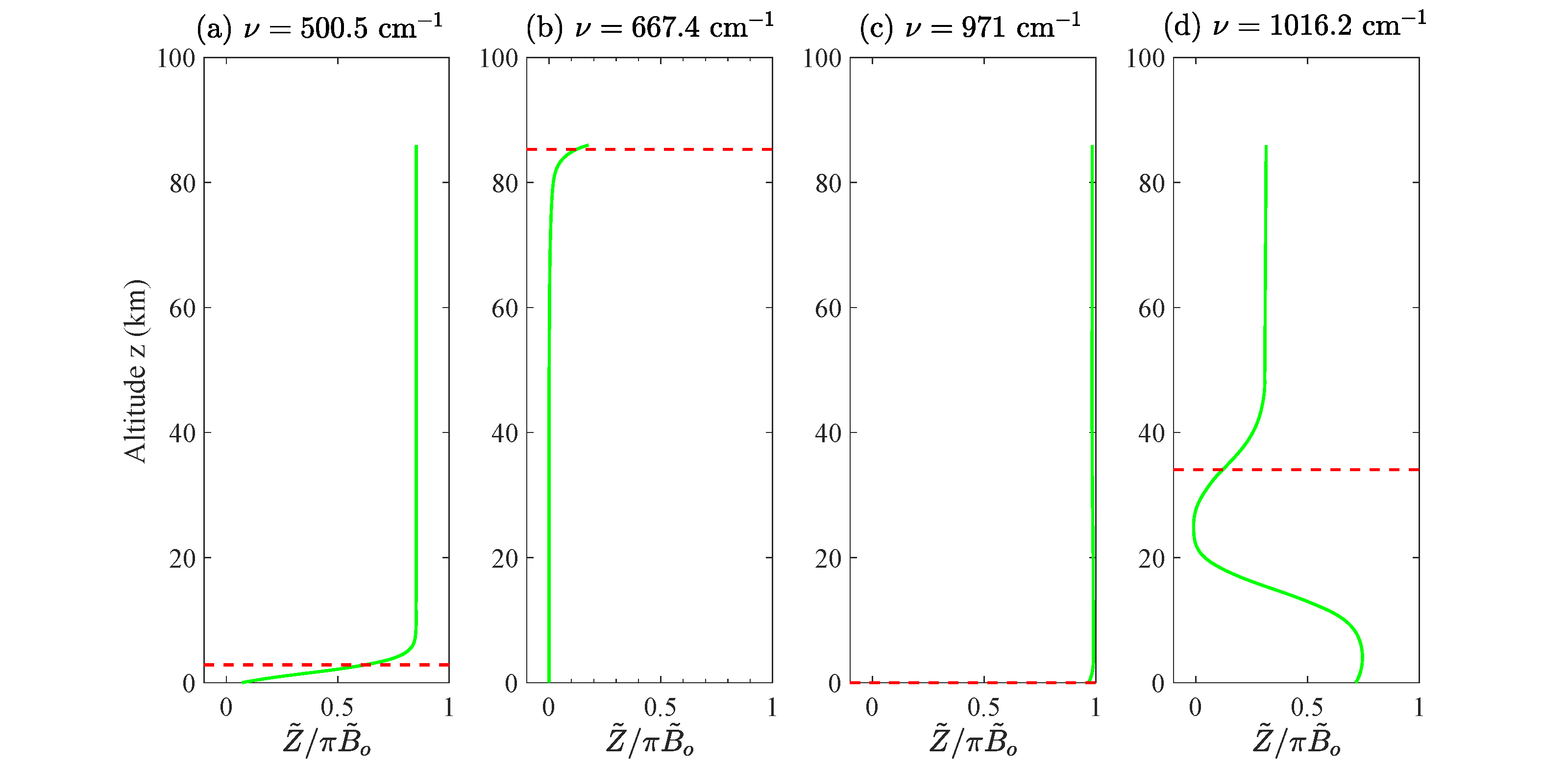}
		\caption{Upward spectral flux $\tilde Z$, for representative frequencies, $\nu$:  (a) $\tau_o = 2.30$, $z_e = 2.9$ km, $\tilde B_o = 134.2$ (i.u.); (b) $\tau_o = 51687$, $z_e = 85.3$ km, $\tilde B_o = 131.8$ (i.u.); (c) $\tau_o = 0.016$, $z_e = 0$ km, $\tilde B_o = 86.4$ (i.u.); (d) $\tau_o = 7.11$, $z_e = 34.0$ km, $\tilde B_o = 79.3$ (i.u.).  Here the spectral intensity unit is 1 i.u. = 1 mW m$^{-2}$ cm sr$^{-1}$.  The dashed red lines mark the emission altitudes.}
		\label{Fig6}
	\end{center}
\end{figure}

\subsection{Clear Skies}

One can accurately calculate thermal radiation transfer for skies without clouds using the Schwarzschild equation

\begin{equation}
\cos \theta \ {{\partial \tilde I} \over {\partial \tau}} = \tilde B - \tilde I.
\label{eqn67}
\end{equation}

\noindent A detailed discussion of how to solve this equation is given in reference \cite{Schwarzschild}, to which we refer readers who are interested in more details. We give only a brief summary of the meaning of (\ref{eqn67}) here.  The monochromatic intensity of radiation with a spatial frequency (wave peaks per cm) $\nu$ and making an angle $\theta$ to the vertical, is  $\tilde I = \tilde I (\nu,\tau,\theta)$.   The vertical optical depth $\tau$ is the number of e-foldings by which vertically propagating surface radiation of frequency $\nu$ is attenuated when it reaches the altitude $z$. Note that in (\ref{eqn67}) the optical depth is measured from the bottom of the atmosphere, but in (\ref{eqn56}) it was measured from the top.  The Planck intensity for isotropic radiation of frequency $\nu$ in thermal equilibrium at an absolute temperature $T$ is 

\begin{equation}
\tilde B(\nu,T) = {{2 h c^2 \nu^3}\over {e^x-1}}.
\label{eqn68}
\end{equation}

\noindent Here $c$ is the speed of light, $h$ is Planck’s constant and the photon energy, in units of $k_B T$ is $x = \nu c h /(k_B T)$.

\begin{figure}[t]
	\begin{center}
		\includegraphics[height=80mm,width=1\columnwidth]{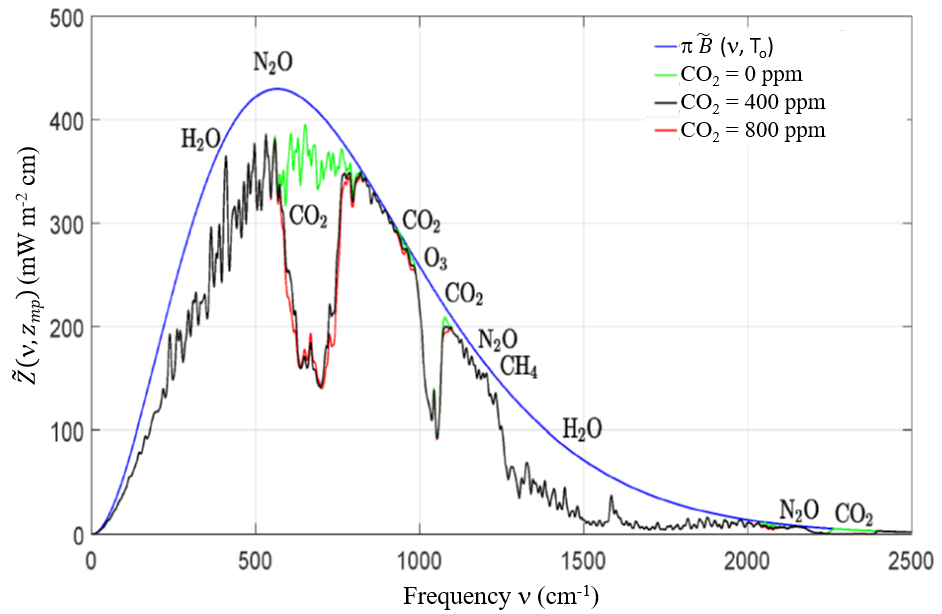}
		\caption{Effects of changing concentrations of carbon dioxide, CO$_2$ on the thermal radiative flux to space from the top of a midlatitude standard atmosphere at altitude $z_{\rm mp} = 86$ km, with the temperature profile of Fig. \ref{Fig5}.  The smooth blue line is the spectral flux from a surface at the temperature $T_0 = 288.7$ K for a transparent atmosphere with no greenhouse gases.  The green line is the flux if all CO$_2$ were removed but with all the other greenhouse gases at their standard concentrations.  The black line is for all greenhouse gases at their standard concentrations.  The red line is for twice the standard concentration of CO$_2$ but with all the other greenhouse gases at their standard concentrations.  Doubling the standard concentration of CO$_2$ from 400 to 800 ppm increases the forcing (the area between the black and red lines) by 3.0 W m$^{-2}$.}
		\label{Fig7}
	\end{center}
\end{figure}

The net upwards flux, $\tilde Z$, the “upwelling” minus “downwelling,” is the sum of the projections of the intensities $\tilde I$ onto the vertical directions

\begin{equation}
\tilde Z(\nu,z) = 2 \pi \int_{-1}^1 \tilde I(\nu,\tau,\theta) \ \cos\theta \ d \cos\theta.
\label{eqn69}
\end{equation}

\noindent Calculations of $\tilde Z$ at various frequencies as a function of altitude are shown in Fig. \ref{Fig6}.  These were done for the temperature profile and greenhouse altitudinal dependences shown in Fig. \ref{Fig5}.  The model included the 5 most important naturally occurring greenhouse gases: H$_2$O, CO$_2$, O$_3$, N$_2$O and CH$_4$.  

For the gray atmosphere, there is a single emission altitude $z_e$, given by (\ref{eqn62}), where most radiation is released to space. The complicated line intensities of Fig. \ref{Fig2} were replaced by a cross section that is independent of the thermal radiation frequencies, and independent of temperature and pressure. All thermal radiation frequencies are attenuated equally for the gray atmosphere. 

For the real atmosphere, Fig. \ref{Fig2} implies an extremely complicated dependence of the emission altitude on frequency, due to the hundreds of thousands of vibrational-rotational lines of the greenhouse molecules. The emission altitude varies dramatically across the thermal radiation spectrum.  Emission altitudes are indicated by dashed-horizontal red lines in Fig. \ref{Fig6}. The green lines show the heat flux carried by frequencies given at the tops of the panels.  

Fig. \ref{Fig6}a shows that for a frequency $\nu = 500.5$ cm$^{-1}$, in the pure rotational band of water vapor, the spectral flux  $\tilde Z$, “breaks out” near the emission altitude $z_e = 2.9$ km, in the lower troposphere. The energy for the flux is radiated by H$_2$O molecules, which extract heat from convecting air parcels.  The atmosphere is cooler at an altitude of 2.9 km than at the surface, so the spectral flux $\tilde Z$ that is radiated to space, the value of the green line at the top of the atmosphere, is only about 80\% of the surface Planck value $\pi \tilde B_o$. 

Fig. \ref{Fig6}b shows that for a frequency $\nu = 667.4$ cm$^{-1}$, in the middle of the strong Q branch of the CO$_2$ bending-mode band, the emission altitude is near the top of the atmosphere, $z_e = 85.3$ km. Here the energy for the emission is not from convecting air parcels. Most of the emitted energy is supplied by the absorption of solar ultraviolet radiation. Heavily absorbed frequencies, like $\nu = 667.4$ cm$^{-1}$, make a negligible contribution to radiative heat transport in the troposphere.

Fig. \ref{Fig6}c shows that for a frequency $\nu = 971$ cm$^{-1}$, in the middle of the atmospheric window, greenhouse gases are so nearly transparent that the surface radiates directly to space. This provides extremely efficient, ballistic heat transfer from the surface to cold space. It is one of the factors that leads to dew or frost on calm, dry, cloud-free nights.  The spectral flux $\tilde Z$ is equal to the spectral Planck flux $\pi \tilde B_o$ at the surface. 

Fig. \ref{Fig6}d shows that for a frequency of $\nu = 1016.2$ cm$^{-1}$, in the middle of the O$_3$ absorption band, the emission altitude $z_e = 34.0$ km is at the same altitude where the O$_3$ concentration is a maximum as is shown in Fig. \ref{Fig5}.  There is little ozone in the troposphere, so the spectral flux $\tilde Z$ in the troposphere is not much smaller than the surface Planck flux $\pi \tilde B_o$. Most of the flux from the surface and troposphere is absorbed by O$_3$ in the lower stratosphere, to be replaced by emission from the warmer, upper parts of the stratosphere, where the temperature maximizes because of absorption of solar ultraviolet light by ozone. The absorbed ultraviolet light provides the energy that is radiated back to space at the frequency $\nu = 1016.2$ cm$^{-1}$.

The effect of the greenhouse gases on the spectral flux $\tilde Z$  emitted at the top of the atmosphere, located at $z_{\rm mp} = 86$ km, is shown in Fig. \ref{Fig7}.  Absorption by pure rotational transitions of H$_2$O at frequencies below 550 cm$^{-1}$, absorption and emission by CO$_2$ near its bending mode frequency of 667 cm$^{-1}$, absorption and emission by O$_3$ at 1016 cm$^{-1}$ and absorption and emission by bending modes of H$_2$O at frequencies above 1200 cm$^{-1}$ dominate the spectrum. Integrating the monochromatic flux over all frequencies gives the total infrared flux

\begin{equation}
Z(z) = \int_0^{\infty} d\nu \ \tilde Z(\nu,z).
\label{eqn70}
\end{equation}

\begin{figure}[t]
\begin{center}
\includegraphics[height=100mm,width=.6\columnwidth]{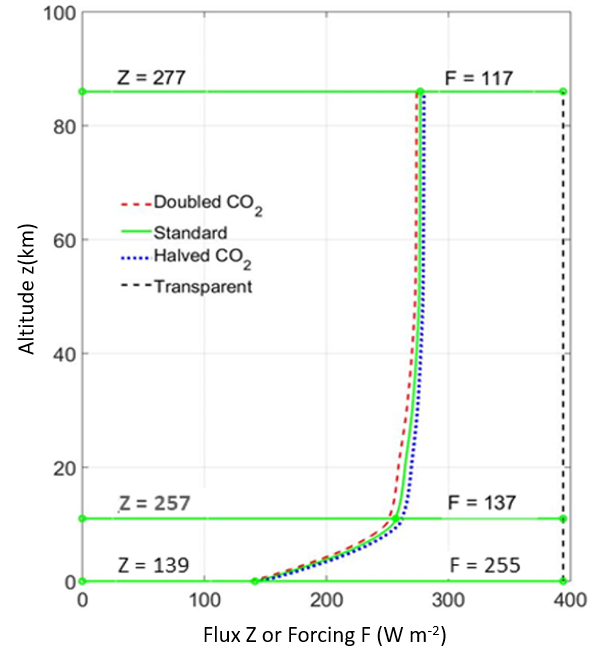}
\caption{Upward flux $Z$ or forcing $F$ calculated from the Schwarzschild equation for a standard atmosphere having 400 ppm CO$_2$, double that amount, 800 ppm, and half that amount, 200 ppm.  The other greenhouse gas concentrations were kept fixed at their standard values shown in Fig. \ref{Fig5}.  Values for $Z$ and $F$ are given at the surface, tropopause altitude of 11 km and top of the atmosphere altitude, $z_{\rm mp} = 86$ km.}
\label{Fig8}
\end{center}
\end{figure}

One can define the forcing as the difference between the infrared flux emitted by the surface at temperature $T_0$ through a transparent atmosphere and the infrared actual flux $Z$ which includes the effects emission and absorption by greenhouse gases

\begin{equation}
F(z) =\sigma {T_0}^4\ –\ Z(z).
\label{eqn71}
\end{equation}

\begin{figure}[t]
	\begin{center}
		\includegraphics[height=100mm,width=.6\columnwidth]{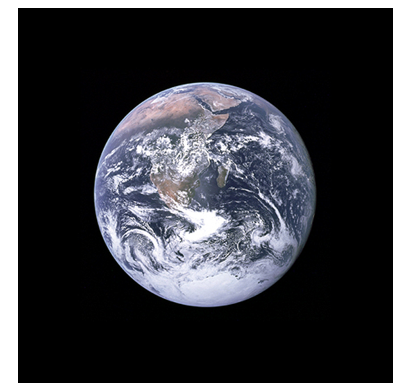}
		\caption{The ``blue marble" a photograph taken by the crew of the Apollo 17 mission to the Moon on December 7, 1972, close to midsummer of the southern hemisphere.  The clouds of the intertropical convergence zone can be seen over Africa, as well as heavier, low cloud cover further south and over Antarctica \cite{NasaMarble}.}
		\label{Fig9}
	\end{center}
\end{figure}

\noindent The upward flux $Z$ and the forcing $F$ are plotted in Fig. \ref{Fig8} as a function of altitude for three different CO$_2$ concentrations while maintaining the other greenhouse gases at their standard concentrations as given in Fig. \ref{Fig5}.  Fig. \ref{Fig8} illustrates the radiative flux, modeled with the Schwarzschild equation (\ref{eqn67}), for the Earth’s real atmosphere. This should be compared to the flux for a gray atmosphere shown in the left panel of Fig. \ref{Fig3}.  Unlike the gray atmosphere, which has only convective heat flux in the troposphere, the real atmosphere has both radiative flux $Z(z)$ at the altitude $z$, which is shown explicitly, and convective flux, which is not shown explicitly, as it was in Fig. \ref{Fig3}.  The convective flux is implicitly given by $Z(11 {\rm \ km}) – Z(z)$, since it vanishes at the tropopause altitude $z = 11$ km.  To the extent that absorption of solar radiation in the troposphere can be neglected, the sum of the steady-state radiative and convective fluxes must be constant to conserve energy.  

Unlike the gray atmosphere of Fig. \ref{Fig3}, where there is negligible radiative flux in the troposphere, Fig. \ref{Fig8} shows that there is a significant amount of radiative flux in the troposphere, even at the surface, because of the atmospheric window, which allows ballistic radiative heat transport to space. 

\begin{figure}[t]
	\begin{center}
		\includegraphics[height=100mm,width=.8\columnwidth]{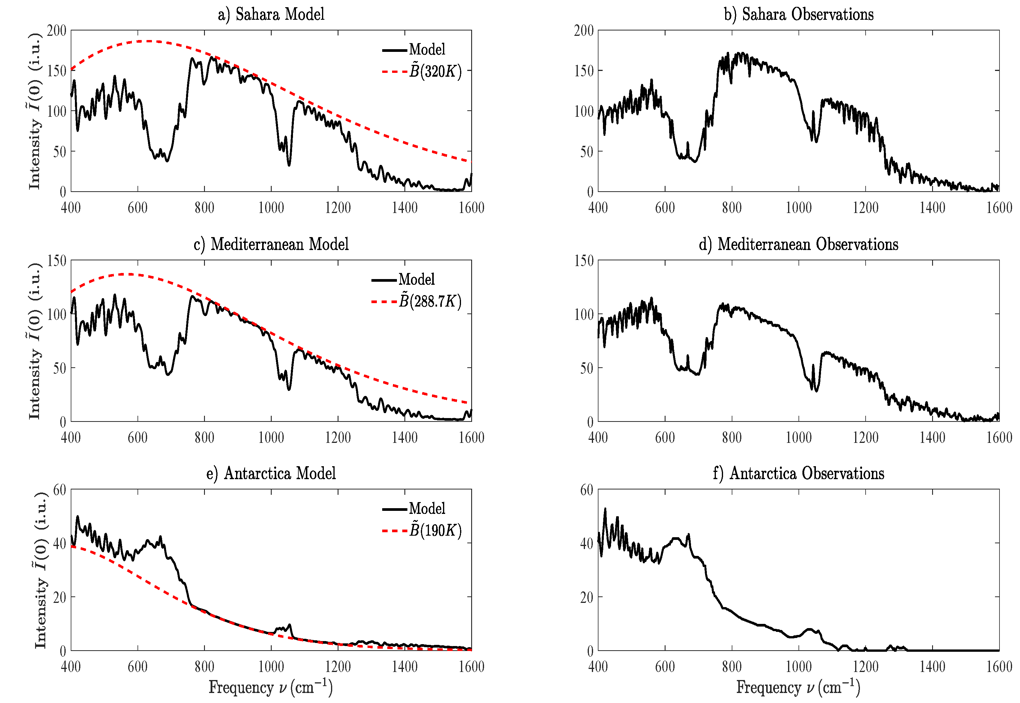}
		\caption{Vertical intensites $I(0)$ at the top of the atmosphere observed with a Michaelson interferometer in a satellite \cite{Hanel}, and modeled with radiation transfer theory for the Sahara desert, Mediterranean and Antarctica.  The intensity unit is 1 i.u. = 1 mW m$^{-1}$ cm sr$^{-1}$.  Radiative forcing is negative over wintertime Antarctica since the relatively warm greenhouse gases in the troposphere mostly CO$_2$, O$_3$ and H$_2$O, radiate more flux to space than the cold ice surface at a temperature of $T=190$ K could radiate through a transparent atmosphere.}
		\label{Fig10}
	\end{center}
\end{figure}

For the gray atmosphere of Fig. \ref{Fig3}, there is an abrupt increase of radiative flux at the tropopause, where all of the heat convected from the surface is released. As shown in Fig. \ref{Fig8}, the varied emission altitudes of the real atmosphere ensure that convected heat continues to be released at all altitudes between the surface and the tropopause. For the real atmosphere, the radiative flux increases roughly linearly, as shown in Fig. \ref{Fig8}, as convective heat is released at increasing altitudes $z$. 

As shown in Fig. \ref{Fig7} and Fig. \ref{Fig8}, the decrease in upwards flux at the top of the atmosphere, $Z(z_{\rm mp})$, due to doubling the CO$_2$ concentration, 3 W m$^{-2}$, is about a 1\% decrease of $Z(z_{\rm mp})$. Doubling the other anthropogenic greenhouse gases, N$_2$O or CH$_4$, results in even smaller changes to the upwards flux. These flux changes are called the instantaneous forcings.  If there were no change in absorbed solar radiation the instantaneous forcing would cause heat to build up in the atmosphere and Earth below. This is where the reliability of climate modeling ends, since no one knows just how the complicated climate system of Earth’s atmosphere and oceans will respond to the small forcings. As shown in reference \cite{WvW1}, if one assumes no change in solar heating, and that convective heat transfer keeps the troposphere approximately adiabatic, the surface temperature will warm by about 1 C from doubling CO$_2$, and the upper stratosphere will cool by about 8 C. 

A surface warming of 1 C is too small a prediction for those who claim that increasing concentrations of greenhouse gases have already led to a climate emergency. Various strongly positive feedbacks have been proposed to predict more surface warming.  One of the favorite feedback mechanisms is lofting water vapor, the most important greenhouse gas to higher, colder emission altitudes. The decreased emission would have to be compensated for by warmer surface temperatures at the bottom of a nearly adiabatic atmosphere, just as for the model gray atmosphere of Fig. \ref{Fig3}.

Like greenhouse gases, emission and absorption of thermal radiation by clouds can modify the release of thermal radiation to space. High clouds with cold tops are very poor thermal radiators. Increases or decreases in high clouds could decrease or increase thermal radiation to space. The iris-effect of Lindzen and Chou \cite{Lindzen} is a possible negative feedback due to changes in high cirrus clouds.  The tops of low clouds are not much cooler than Earth’s surface, so low clouds decrease thermal radiation to space much less than high clouds.  

Thick low clouds also reflect sunlight back to space before it is absorbed and converted to atmospheric heat. Unlike greenhouse gases, which are nearly transparent to sunlight, clouds can be very non-transparent due to scattering and weak absorption of visible light. The beautiful images of Earth taken from space, “the blue marble”, show the large amount of sunlight reflected from clouds, land and oceans \cite{NasaMarble}.  Any sunlight you can see has not heated the Earth. Averaged over its surface, Earth reflects about 30\% of sunlight back to space; in other words, the “Bond albedo” of Earth is about 0.3.

Another positive feedback that has been proposed is a change in the albedo in polar regions due to melting ice.  However, a reduction in summer Arctic ice of about 2 million km$^2$ as has been observed during primarily from 1990 to 2007 \cite{NSIDC} represents less than 0.4\% of the Earth’s surface and therefore has primarily a regional as opposed to a global impact.

An important check on any calculation is to compare observations to modeled results.  Fig. \ref{Fig10} shows intensities emitted by the top of the atmosphere as measured by an interferometer in a satellite.   The modeled results were obtained by solving the Schwarzschild equation and using the appropriate temperature profiles.  The surface water vapor concentration varied significantly from 31,000 ppm for the Sahara, 12,000 ppm for the Mediterranean and only 2,000 ppm for Antarctica.  The excellent quantitative agreement with observations at very different latitudes give confidence that the calculations of infrared flux are correct. 

\section{Summary}

Greenhouse gases are responsible for the most striking feature of Earth’s atmosphere, a lower troposphere and an upper stratosphere. In the troposphere, below the tropopause boundary, a large fraction of the energy flux from the solar heated surface is carried by convection, and not by thermal radiation.  Convection maintains average temperature lapse rates in the troposphere that are close to adiabatic. In the stratosphere, most of the upward heat flux is carried by radiation. Greenhouse gases warm the surface because they increase the “thermal resistance” of the atmosphere to the vertical flow of energy from the solar-heated surface to space.  The larger the thermal resistance between the surface and the emission altitude, the larger the temperature difference needed to drive the solar energy absorbed by the surface back to space. Without the thermal resistance induced by greenhouse gases, Earth’s surface would be much colder and life as we know it would not be possible.

Increasing carbon dioxide will cause a small additional surface warming.  It is difficult to calculate exactly how much, but our best estimate is that it is about 1 C for every doubling of CO$_2$ concentration, when all feedbacks are correctly accounted for. Alarming predictions of dangerous warming require large positive feedbacks. The most commonly invoked feedback is an increase in the concentration of water vapor in the upper troposphere.  But most climate models have predicted much more warming than has been observed, so there is no observational support for strong positive feedbacks \cite{Christy,Fyfe}. Indeed, most feedbacks in nature are negative as expressed by Le Chatelier’s Principle \cite{Chatelier}:
{\it When any system at equilibrium for a long period of time is subjected to a change in concentration, temperature, volume or pressure, the system changes to a new equilibrium, and this change partly counteracts the applied change.}

We have barely touched atmospheric dynamics, perhaps the most interesting part of the grand drama of weather and climate.  We are all familiar with manifestations of atmospheric dynamics:  warm fronts, cold fronts, droughts, floods, hurricanes, tornados etc. Equally fascinating ocean dynamics, like the El Nino cycles of the tropical Pacific Ocean, also contributes to weather and climate.  Earth’s atmosphere works like an extremely complicated engine that transforms heat from the Sun into the work that drives the winds, the weather and ocean dynamics. Greenhouse gases are the heat exchanger which allows the atmospheric heat engine to dump waste heat into cold space.

\section*{Acknowledgements}
 The Canadian Natural Science and Engineering Research  Council provided financial support of one of us.

\end{document}